\documentclass{article}

\newlength{\defaultparindent}
\begin{document}
\begin{center}
\LARGE{\bf{Time Symmetry in Microphysics}}\footnote{This is a 
draft of a paper for a symposium at 
the Biennial Meeting of the Philosophy of Science Association, 
Cleveland, Ohio, 1--3 November, 1996. I am grateful 
for comments on previous versions from audiences 
at ANU, UWO and
Columbia in 1995, and 
at the AAHPSSS Conference at the University of Melbourne in June, 
1996.}
\bigskip

\large{Huw Price}%
\footnote{School of Philosophy, University of Sydney, Australia 
2006 [\texttt{huw@mail.usyd.edu.au}].}%

\end{center}\bigskip\bigskip

\noindent%
{\sc Abstract:} Physics takes for granted that interacting 
physical systems with no common history are independent, before 
their interaction. This principle is time-asymmetric, for no 
such restriction applies to systems with no common future, after 
an interaction. The time-asymmetry is normally attributed to 
boundary conditions. I argue that there are two distinct independence 
principles of this kind at work in contemporary physics, one 
of which cannot be attributed to boundary conditions, and therefore 
conflicts with the assumed T (or CPT) symmetry of microphysics. 
I note that this may have interesting ramifications in quantum 
mechanics.
\bigskip

\section{Introduction}

\setlength{\parskip}{0pt}Consider a photon, passing through 
a polariser. According to the standard model of quantum mechanics, 
the state of the photon after the interaction reflects the orientation 
of the polariser. Not so before the interaction, of course: in 
quantum mechanics, as elsewhere in physics, we don't expect %
{\it preinteractive} correlations.

Writers who notice this time 
asymmetry---postinteractive correlations, but no preinteractive 
correlations---sometimes see it as an objection to the standard 
model of quantum mechanics. To most, however, it seems hardly 
worthy of notice. True, the asymmetry may be a little puzzling, 
but its individual components---that interactions may establish 
correlations, and that there are no preinteractive correlations---seem 
plausible enough. If we were to try for symmetry, which should 
we give up? Besides, the principle that there are no preinteractive 
correlations plays an important role elsewhere in the physics 
of time-asymmetry, where there is a well-established view to 
the effect that it is not in conflict with the T-symmetry of 
underlying physical laws. Thus there seems to be a precedent 
for the asymmetry we find in quantum mechanics, and no reason, 
on reflection, to doubt our initial intuitions.

I think the calm is illusory, however, and my aim here is to 
reveal the troubled waters beneath these rather slippery intuitions. 
I shall argue that the time asymmetry embodied by the standard 
model is quite distinct from its supposed analog elsewhere in 
physics, and cannot be reconciled with the T-symmetry of the 
laws of physics in the same way. Given T-symmetry, I contend, 
pre- and postinteractive correlations should be on the same footing 
in microphysics. Any reason for objecting to preinteractive correlations 
is a reason for objecting to postinteractive correlations, and 
any reason for postulating postinteractive correlations is a 
reason for postulating preinteractive correlations. 

I emphasise that for the bulk of the paper, the link with quantum 
mechanics is indirect. The standard model provides vivid examples 
of the intuitions I want to examine, but my interest is in the 
intuitions themselves, not in the quantum mechanical examples. 
However, I close with a comment on the significance of my argument 
for the puzzles of quantum mechanics. Briefly, its effect seems 
to be to undermine a crucial presupposition of the standard arguments 
that quantum mechanics cannot be interpreted in more-or-less 
classical terms.

\section{Two kinds of preinteractive independence}

\setlength{\parskip}{0pt}The principle that there are no preinteractive 
correlations has famous connections with the most striking time-asymmetry 
in physics, that of the second law of thermodynamics. The connections 
emerge at two levels. First, Boltzmann's %
{\it H}-Theorem derives its time-asymmetry from an assumption 
concerning the preinteractive independence of interacting microsystems. 
(This assumption needs to be time-asymmetric, of course, since 
otherwise the theorem would apply equally in either temporal 
direction.)

At a more intuitive level, familiar 
low-entropy systems are associated with striking postinteractive 
correlations. To make this point vivid, think of the astounding 
preinteractive correlations we observe if we view ordinary processes 
in reverse. Think of the tiny droplets of champagne, forming 
a pressurised column and rushing into a bottle, narrowly escaping 
the incoming cork. Or think of the countless (genuine!) fragments 
of the True Cross, making their precisely choreographed journeys 
to Jerusalem. Astounding as these feats seem, they are nothing 
but the mundane events of ordinary life, viewed from an unfamiliar 
angle. Correlations of this kind are ubiquitous in one temporal 
sense---when they occur %
{\it after} some central event, from our usual perspective---but 
unknown and incredible in the other temporal sense.

In the macroscopic world of ordinary experience, then, the presence 
of postinteractive correlations and the absence of preinteractive 
correlations is closely associated with the thermodynamic asymmetry. 
It is an old puzzle as to where this asymmetry comes from, and 
especially as to how it is to be reconciled with the apparent 
T-symmetry of the underlying laws of physics. The orthodox view 
is that the asymmetry of thermodynamics is a matter of boundary 
conditions: factlike rather than lawlike, as physicists often 
say. The contemporary version of this view traces the low-entropy 
history of familiar physical systems to the condition of the 
early universe. True, many hope that this early condition will 
itself be explicable as a natural consequence of cosmological 
laws, in which case the resulting asymmetry is not strictly factlike. 
Nevertheless, the success of this program would preserve the 
intuitive distinction between the symmetry of %
{\it local} dynamical laws, and the asymmetry of the boundary 
conditions supplied to these symmetric laws in typical real systems.

It seems to be assumed that the kind of asymmetry exemplified 
by photons and polarisers can be accommodated within this general 
picture, but I want to show that this is not so. If there is 
an asymmetry in microphysics of this kind, it cannot be accorded 
the status of a (locally) factlike product of boundary conditions. 
This is because, unlike in the thermodynamic case, there is no 
observational evidence for the required asymmetry in boundary 
conditions. On the contrary, our sole grounds for thinking that 
the boundary conditions are asymmetric in the relevant sense 
is that we already take for granted the principle that there 
are post- but not preinteractive correlations of the relevant 
kind. In effect, then, this principle operates in a lawlike manner, 
in conflict with the assumed T-symmetry of (local) dynamical 
laws.

The first step is to show that the kind of postinteractive correlation 
displayed by the photon is quite distinct from that associated 
with low-entropy systems, such as the champagne bottle. With 
a little thought, this distinction is easy to draw. For one thing, 
the correlations associated with low-entropy systems are essentially 
``communal'', in the sense that they involve correlations among 
the behaviour of very large numbers of individual systems. But 
the photon correlations are individualistic, in the sense that they involve the 
simplest kinds of interactions between one entity and another.

Second, the photon case is not dependent on the thermodynamic 
history of the system comprising the photon and the polariser, 
or any larger system of which it might form a part. Imagine a 
sealed black box containing a rotating polariser, and suppose 
that the thermal radiation inside the box has always been in 
equilibrium with the walls. We still expect the photons comprising 
this radiation to establish the usual postinteractive correlations 
with the orientation of the polariser, whenever they happen to 
pass through it. The presence of these postinteractive correlations 
does not require that entropy was lower in the past. By symmetry, 
then, the absence of matching preinteractive correlations cannot 
be deduced---at any rate, not directly---from the fact that entropy 
does not decrease toward the future: a world in which photons 
were correlated with polarisers before they interacted would 
not necessarily be a world in which the second law of thermodynamics 
did not hold.

It will be helpful to have labels for the two kinds of preinteractive 
independence just distinguished. I'll call the principle that 
there are no entropy-reducing correlations ``%
{\it H}-Independence'', in light of its role in the %
{\it H}-Theorem, and the principle that there are no preinteractive 
correlations between individual micro-systems ``micro-independence'' 
(``%
{\normalsize $\mu$}Independence'', for short).

\section{Initial randomness?}

I have argued that observational evidence 
for %
{\it H}-Independence need not be observational evidence for 
{\normalsize $\mu$}Independence---at any rate, not directly. 
There might be an indirect argument in the offing, however. Perhaps 
the second law supports some hypothesis about the initial conditions 
of the universe, an independent consequence of which is that 
photons are not correlated with polarisers before they interact. 
For example, it is often suggested that the explanation for the 
second law lies in the fact that the initial microstate of the 
universe is as random as it can be, given its low-entropy macrostate. 
Wouldn't this hypothesis also explain why photons are not correlated 
with future polarisers?

In my view this hypothesis is 
independently unsatisfactory. In particular, it is doubtful whether 
the required boundary condition can be specified in a nonvacuous 
way---i.e., other than as the condition that the initial state 
of the universe is such that the second law holds. (See Price 
1996, 42.) General defects to one side, however, the hypothesis 
turns out to be irrelevant to the issue at hand. In effect, the 
suggestion is that if systems comprising photons and polarisers 
are allowed a free choice of the available initial microstates, 
there can be no general correlation between the states of incoming 
photon-polariser pairs. If this were true, what would it mean 
for the ordinary postinteractive correlations? Do these require 
that the final conditions be less than completely random? Not 
if we understand the choice to be made from those situations 
permitted by the relevant physical laws---in other words, from 
the phase space of the system in question. Of course, if we think 
of nature making its choice from some larger set of possibilities, 
then the laws themselves constitute restrictions on the available 
options. Only choices in accordance with the laws are allowed. 
But a random choice from phase space (or, equivalently, from 
the set of trajectories of a deterministic system) is by definition 
a choice from among (all and only) the options allowed by the 
laws.

Thus lawlike postinteractive correlations are not incompatible 
with randomness of final conditions. By symmetry, matching preinteractive 
correlations would not require non-random initial conditions. 
Hence %
{\normalsize $\mu$}Independence receives no support from the 
hypothesis that initial randomness explains the thermodynamic 
asymmetry.

\section{Colliding beams?}

There is another argument in the literature 
to the effect that there is indirect observational evidence for 
{\normalsize $\mu$}Independence. It turns on the idea that by 
postulating %
{\normalsize $\mu$}Independence, we are able to explain certain 
observable phenomena. I think this argument is due originally 
to O. Penrose and R. Percival (1962), who formulate a principle 
of preinteractive independence they call the Law of Conditional 
Independence. As their terminology indicates, Penrose and Percival 
take this to be a lawlike principle. In favour of this view, 
they argue that the principle is able to explain a variety of 
otherwise inexplicable irreversible processes. 

The claim that Conditional Independence 
is lawlike has not been widely accepted, but it does seem a common 
view in physics that Penrose and Percival's examples provide 
indirect observational evidence for preinteractive independence. 
A typical example concerns the scattering which occurs when two 
tightly organised beams of particles are allowed to intersect. 
The argument is that this scattering is explicable if we assume 
that there are no prior correlations between colliding pairs 
of particles (one from each beam)---and hence that the scattering 
pattern reveals the underlying independence of the motions of 
the incoming particles.

In fact, however, %
{\normalsize $\mu$}Independence is neither necessary nor sufficient 
here. The explanation rests entirely on the absence of entropy-reducing 
correlations between the incoming beams---i.e. on %
{\it H}-Independence---not on %
{\normalsize $\mu$}Independence at the level of individual particle 
pairs. In other words, the asymmetry involved in these cases 
is nothing more than the familiar thermodynamic asymmetry, from 
which---as we have seen---%
{\normalsize $\mu$}Independence is supposed to be distinct.

I'll offer short and long arguments for this conclusion. The 
short argument simply appeals to cases in which it seems intuitively 
clear that there is no microscopic asymmetry---Newtonian particles, 
for example. In these cases there seems to be nothing to sustain 
any asymmetry at the level of individual interactions, and yet 
we still expect colliding beams to scatter. This suggests that 
the scattering is associated with the lack of some global correlation, 
not with anything true of individual particle pairs. 

The longer argument goes like this. We suppose that there is 
a microscopic asymmetry of %
{\normalsize $\mu$}Independence, distinct from the correlations 
associated with the thermodynamic asymmetry, and yet compatible 
with the T-symmetry of the relevant dynamical laws. We then construct 
a temporal inverse of the scattering beam experiment, and show 
that it displays (reverse) scattering, despite the assumed absence 
of the postinteractive analog of %
{\normalsize $\mu$}Independence. By symmetry, this shows that 
{\normalsize $\mu$}Independence is not necessary to explain the 
scattering observed in the usual case. Finally, a variant of 
this argument shows that %
{\normalsize $\mu$}Independence is also insufficient for the 
scattering observed in the usual case. 

If %
{\normalsize $\mu$}Independence were necessary for scattering, 
in other words, then scattering would not occur if the experiment 
were run in reverse. It is difficult to replicate the experiment 
in reverse, for we don't have direct control of final conditions. 
But we can do it by selecting the small number of cases which 
do satisfy the desired final conditions from a larger sample. 
We consider a large system of interacting particles of the kind 
concerned, and consider only those pairs of particles which emerge 
on two tightly constrained trajectories (one particle on each), 
having perhaps interacted in a specified region at the intersection 
of these two trajectories (though not with any particle which 
does not itself emerge on one of these trajectories). We then 
consider the distribution of initial trajectories, before interaction, 
for these particles. What is the most likely distribution? If 
the dynamical laws are T-symmetric, then it must be simply the 
distribution which mirrors the predicted scattering in the usual 
case.

The argument can be made more explicit by describing a symmetric 
arrangement, subsets of which duplicate both versions of the 
experiment. Consider a spherical shell, divided by a vertical 
plane. On the inner face of the left hemisphere is an arrangement 
of particle emitters, each of which produces particles of random 
speed and timing, directed towards the centre of the sphere. 
In the right hemisphere is a matching array of particle detectors. 
Dynamical T-symmetry implies that if the choice of initial conditions 
is random, the global history of the device is also time-symmetric: 
any particular pair of particle trajectories is equally likely 
to occur in its mirror-image form, with the position of emission 
and absorption reversed.

We can replicate the original collimated beam experiment by choosing 
the subset of the global history of the device containing particles 
emitted from two chosen small regions on the left side. Similarly, 
we can replicate the reverse collimated beam experiment by choosing 
the subset of the history of the entire device containing particles 
absorbed at two chosen small regions on the right side. In the 
latter case, the particles concerned will in general have been 
emitted from many different places on the left side. This follows 
from the fact that the initial conditions are a random as possible, 
compatible with the chosen final conditions. Thus we have scattering 
in the initial conditions, despite the assumed lack of postinteractive 
{\normalsize $\mu$}Independence between interacting particles.

Thus if there were postinteractive correlations of the kind denied 
to the preinteractive case by %
{\normalsize $\mu$}Independence, they would not stand in the 
way of scattering in the reverse experiment---scattering in that 
case is guaranteed by the assumption that the  initial conditions 
are as random as possible, given the final constraints. By symmetry, 
however, this implies that %
{\normalsize $\mu$}Independence is not necessary to produce scattering 
in the normal case. We would have scattering without %
{\normalsize $\mu$}Independence, provided that the choice of 
trajectories is as random as possible, given the initial constraints. 
(Don't suggest that this is the same thing as %
{\normalsize $\mu$}Independence. If that were true, %
{\normalsize $\mu$}Independence would not fail in the postinteractive 
case, and there not be the assumed microscopic asymmetry.)

A third version of the experiment can be used to show that %
{\normalsize $\mu$}Independence is not sufficient to explain 
what happens in the normal case. Assume %
{\normalsize $\mu$}Independ\-ence again, and consider the subset 
of the first experiment in which we have collimation on the right, 
as well as the left---in other words, in which we impose a final 
condition, as well as an initial condition. In this case, we 
have no scattering, despite %
{\normalsize $\mu$}Independence. Again, it is no use saying that 
the imposition of the final condition amounts to a denial of 
{\normalsize $\mu$}Independence: if that were true, the asymmetry 
of %
{\normalsize $\mu$}Independence in the normal case would amount 
to nothing more than the presence of a low-entropy initial condition, 
in conflict with the supposition that %
{\normalsize $\mu$}Independence differs from %
{\it H}-Independence.

In other words, %
{\normalsize $\mu$}Independence is both insufficient and unnecessary 
to explain the phenomena observed in these scattering experiments. 
The differences between the various versions of the experiment 
are fully explained by the different choices of initial and final 
boundary conditions. The asymmetry of the original case stems 
from the fact that we have a low-entropy initial condition (consisting 
in the fact that the beam are initially collimated) but no corresponding 
final condition. The issue as to why this is the case that occurs 
in nature is a sub-issue of that of the origins of the thermodynamic 
asymmetry in general. It has nothing to do with any further asymmetry 
of kind described by %
{\normalsize $\mu$}Independence.

\section{What to do about $\mu$Independence}

It seems that as it currently operates 
in physics, then, %
{\normalsize $\mu$}Independence is not an a posteriori principle 
derived from observation, but a lawlike principle in its own 
right. We don't %
{\it observe }that the incoming photon is not correlated with 
polariser through which it is about to pass. Rather, we rely 
on a tacit meta-law that laws enforcing preinteractive correlations 
would be unacceptable. In a sense, then, we do take it for granted 
that there is an asymmetry in the boundary conditions of the 
kind required by %
{\normalsize $\mu$}Independence: not because we have empirical 
evidence for such an asymmetry, however, but only because we 
have framed the laws in the light of %
{\normalsize $\mu$}Independence. We allow dynamical principles 
producing postinteractive correlations, while disallowing their 
preinteractive twins.

Conceding that %
{\normalsize $\mu$}Independence is lawlike does not improve its 
prospects, of course; it simply 'fesses-up to the principle's 
current role in microphysics. In one important sense it makes 
its prospects very much worse, for as a lawlike principle, %
{\normalsize $\mu$}Independence conflicts with T-symmetry. We 
might be justified in countenancing such a conflict if there 
were strong empirical evidence for a time-asymmetric law, but 
the supposed evidence for %
{\normalsize $\mu$}Independence turns out to rely on a different 
asymmetry altogether.

What are the options at this point? First, we might look for 
other ways of defending %
{\normalsize $\mu$}Independence. Unless this evidence is a posteriori, 
however, its effect will be simply to deepen the puzzle about 
the T-asymmetry of microphysics. Moreover, although there is 
undoubtedly more to be said about the intuitive plausibility 
of %
{\normalsize $\mu$}Independence, I suspect that the effect of 
further investigation is to explain but not to %
{\it justify} our intuitions. For example, the intuitive appeal 
of %
{\normalsize $\mu$}Independence may rest in part on a feature 
of human experience, the fact that in practice our knowledge 
of things in the physical world is always postinteractive, not 
preinteractive. The exact explanation of this asymmetry is rather 
tricky. It seems to depend in part on our own time-asymmetry 
as structures in spacetime, and in part on broader environmental 
aspects of the general thermodynamic asymmetry. Whatever its 
exact provenance, however, it seems to provide no valid grounds 
for extending the intuitions concerned to microphysics. 

Similarly, as I've argued elsewhere (1996, 181--4), some apparent 
postinteractive dependencies turn out to be associated with a 
temporal asymmetry in counterfactual reasoning---roughly, the 
fact that we ``hold fixed'' the past, when considering the consequences 
of counterfactual conditions. Given a conventional account of 
this aspect of counterfactual reasoning, the asymmetries concerned 
are thus demystified, in the sense that they are shown to require 
no independent asymmetry in the physical systems concerned. Again, 
some of the intuitive appeal of %
{\normalsize $\mu$}Independence is thereby accounted for, but 
in a way which does nothing to clarify the puzzle of the photon 
case. 

Another response to the puzzle would be to try to restore T-symmetry 
in microphysics by excising postinteractive correlations, rather 
than by admitting preinteractive correlations. The standard model 
of quantum mechanics might be first in line, for example. The 
surgery required is likely to be rather radical, however. Without 
postinteractive correlation of some sort, how is it possible 
for a measuring device to record information about an object 
system? That aside, the move seems misguided. It does nothing 
to justify %
{\normalsize $\mu$}Independence, and restores symmetry by creating 
two puzzles where previously we had one. 

In my view, the only option which really faces up to the problem 
is that of admitting that our intuitions might be wrong, and 
that %
{\normalsize $\mu$}Independence might indeed fail in microphysics. 
I want to finish with a few remarks on the possible relevance 
of this option in quantum mechanics. In order to clarify the 
force of these remarks, I emphasise again that up to this point, 
my references to quantum mechanics have been somewhat inessential. 
The standard model of quantum mechanics provides the most vivid 
examples of an asymmetry we find it easy to take for granted 
in microphysics, but the case against this asymmetry has been 
essentially classical. The main point is that despite common 
opinion to the contrary, it is not associated with the classical 
asymmetry of thermodynamics. In effect, then, the case against 
{\normalsize $\mu$}Independence constitutes a prior constraint 
on the interpretation of quantum mechanics.

\section{$\mu$Independence and quantum mechanics}

Surprisingly, %
{\normalsize $\mu$}Independence turns out to be a fundamental 
assumption of the main arguments taken to show that the quantum 
world is puzzlingly nonclassical. In particular, Bell's Theorem 
depends on the assumption that the state of an object system 
is independent of the setting of a measurement device, prior 
to their interaction. Thanks to %
{\normalsize $\mu$}Independence, this independence assumption 
has often seemed so uncontentious as to pass without comment. 
Bell himself considered relaxing it, but even he tended to think 
about this possibility in a way which doesn't conflict with %
{\normalsize $\mu$}Independence. (His suggestion, which he called 
``superdeterminism'', was that the correlation might established 
by an additional common cause in the past, not simply in virtue 
of the existing interaction in the future; see Bell et. al., 
1985.) 

More recent arguments for nonlocality 
(the GHZ cases; see e.g., Clifton, Pagonis and Pitowsky 1992) 
also depend on this independence assumption. Without %
{\normalsize $\mu$}Independence, then, there seems to be no firm 
reason to think that quantum mechanics commits us to nonlocality. 
Many commentators have noted that in principle, the Bell correlations 
are easily explicable if hidden common causes may lie in the 
future, as well as in the past. My point is that if %
{\normalsize $\mu$}Independence is rejected on classical grounds, 
this is precisely what we should expect.

There is a similar impact on the no hidden variable theorems 
(e.g. Kochen and Specker 1967), which argue that no system of 
pre-existing properties could reproduce the predictions of quantum 
mechanics, at least in certain cases. %
{\normalsize $\mu$}Independence serves to justify the assumption 
that a single hidden state must reproduce the quantum predictions 
for any %
{\it possible} next measurement. If the hidden state is allowed 
to vary with the nature of the measurement, the problem is relatively 
trivial. (In Bohm's 1952 hidden variable theory, the trick is 
to allow measurement to have an instantaneous effect on the hidden 
variables; again, however, %
{\normalsize $\mu$}Independence underpins the assumption that 
the effect must be instantaneous, rather than advanced.) Abandoning 
{\normalsize $\mu$}Independence might thus resuscitate the hidden 
variable approach, and with it an old solution to the measurement 
problem: If collapse corresponds merely to a change in information, 
it is unproblematic. 

Thus %
{\normalsize  $\mu$}Independence plays a crucial role in the 
main arguments taken to show that quantum mechanics has puzzling 
nonclassical consequences. Imagine how things would have looked 
if physics had considered abandoning %
{\normalsize $\mu$}Independence on symmetry grounds, before the 
development of quantum mechanics. Quantum mechanics would then 
have seemed to provide an additional argument against %
{\normalsize $\mu$}Independence, by reductio: given quantum mechanics, 
{\normalsize $\mu$}Independence implies such absurdities such 
as nonlocality and the measurement problem. Against this background, 
then, experimental confirmation of the Bell correlations would 
have seemed to provide empirical data for which the best explanation 
is that %
{\normalsize $\mu$}Independence does fail, as already predicted 
on symmetry grounds.

Of course, from a contemporary standpoint it is difficult to 
see things in these terms. Leaving aside our intuitive commitment 
to %
{\normalsize $\mu$}Independence, the quantum puzzles have lost 
much of their capacity to shock---familiarity has bred a measure 
of contentment in physics, and the imagined reductio has lost 
its absurdum. Regaining a classical perspective would not be 
an easy step, or one to be attempted lightly, but it does seem 
worth entertaining. By abandoning a habit of thought which already 
seems to conflict with well-established principles of symmetry, 
we might free quantum mechanics of consequences which once seemed 
intolerable in physics, and might do so again. 

\newpage

\section{References}

\begin{list}{ }{%
\setlength{\leftmargin}{0pt}\setlength{\rightmargin}{0pt}%
\setlength{\topsep}{0pt}\setlength{\partopsep}{0pt}}
\item {\sc Bell, 
J., Clauser, J., Horne, M. and Shimony, A.} (1985), ``An Exchange 
on Local Beables'', %
{\it Dialectica} %
{\it 39:} 86-110.

\item {\sc Bohm, D.} (1952), ``A Suggested Interpretation of Quantum Theory 
in Terms of Hidden Variables'', %
{\it Physical Review} %
{\it 85:} 166--193.

\item {\sc Clifton, 
R., Pagonis, C. and Pitowsky, I.} (1992), ``Relativity, Quantum 
Mechanics and EPR'', in D. Hull, M. Forbes and K. Okruhlik, (eds.), 
{\it PSA 1992, Volume 1}, Chicago: Philosophy of Science Association, 
pp. 114-28.

\item {\sc Kochen, 
S. and Specker, E. P.} (1967), ``The Problem of Hidden Variables 
in Quantum Mechanics'', %
{\it Journal of Mathematics and Mechanics} %
{\it 17:} 59--87 .

\item {\sc Penrose, O. and Percival, I. C.} (1962), ``The Direction of Time'', 
{\it Proceedings of the Physical Society} %
{\it 79:} 605--616.

\item {\sc Price, H}. (1996), %
{\it Time's Arrow and Archimedes' Point}, New York: Oxford University 
Press.

\end{list}\end{document}